\input{aipcheck}

\documentclass[
    ,final            
  ]
  {aipproc}

\layoutstyle{8x11single}

\begin{document}

\title{Ion-beam therapy: from electron production in tissue like media to DNA damage estimations}

\classification{61.80.-x; 87.53.-j; 41.75.Aki; 34.50.Bw 
                }
\keywords      {Physical radiation effects, radiation damage; Effects of ionizing radiation on biological systems; Positive-ion beams; Energy loss and stopping power}

\author{Emanuele Scifoni}{
  address={Frankfurt Institute for Advanced Studies, Ruth Moufang Str. 1, 60438 Frankfurt am Main, Germany}
}

\author{Eugene Surdutovich}{
  address={Department of Physics, Oakland University,Rochester, Michigan 48309, USA}
}

\author{Andrey Solov'yov}{
  address={Frankfurt Institute for Advanced Studies, Ruth Moufang Str. 1, 60438 Frankfurt am Main, Germany}
}

\author{Igor Pshenichnov}{
  address={Frankfurt Institute for Advanced Studies, Ruth Moufang Str. 1, 60438 Frankfurt am Main, Germany} ,altaddress={Institute for Nuclear Research, Russian Academy of Science, 117312
  Moscow, Russia}
}
\author{Igor Mishustin}{
  address={Frankfurt Institute for Advanced Studies, Ruth Moufang Str. 1, 60438 Frankfurt am Main, Germany} ,altaddress={Kurchatov Institute, Russian Research Center, 123182
  Moscow, Russia}
}
\author{Walter Greiner}{
  address={Frankfurt Institute for Advanced Studies, Ruth Moufang Str. 1, 60438 Frankfurt am Main, Germany}
}

\begin{abstract}
Radiation damage induced by ion beams is traditionally treated at different levels of theoretical approaches, for the different scales and mechanisms involved. 
We present here details of a combined approach that, from a method at a nanoscopic scale, attempts to merge with higher scales existing results, by ``tuning'' the analytical method employed when extended to larger scale and so yielding a consistent picture of the entire process. Results will show the possibility to get a good agreement with macroscale methods and, on the other hand, to produce a reliable electron energy spectra to be used for DNA damage estimations.
\end{abstract}

\maketitle

\section{Introduction}
Presently, ion-beam cancer therapy, after a number of outstanding clinical results, is a widely used treatment tool based on modern theoretical and computational methods of basic science \cite{Kraft05}. Indeed, while the general principles governing the phenomenon are well understood for decades, many details of the involved physical processes, that lead from the incidence of an energetic ion in the media to the biological damage, are still far from being explained and quantitatively estimated at a nanoscopic level. More, between the many different approaches exploited at different scales, with a number of very accurate works \cite{Goodhead06,Nikjoo99,Nikjoo06}, there is no unified picture of the entire mechanism .
A new approach is presented for a preliminary estimation of the damage occurring in a DNA unit placed at a given distance from a $^{12}$C ion track (see also in this proceedings book \cite{ourRADAM1}).
 A major part of the damaging effects following the passage of
an energetic ion in a tissue like medium, are connected, directly or indirectly, to the number
and energy distributions of secondary electrons produced by the primary ionization of the
medium molecules . Our method starts from a calculation of the energy  spectrum of the electrons originally produced by the ion beam [3,4], followed by the evaluation of their possible pathways from the track. This energy spectrum is obtained from the analytical calculation of the singly differentiated cross section of ionization of the medium. After integration  this yields the penetration depth and the typical profile of the Linear Energy Transfer (LET).
The key idea is then  to set up the equation for  electron production , and then integrate it to large scale for checking consistency, with macroscopic scale results.
In other words, by looking at the macroscopic quantities one can adjust the model for electron production the analytical method in order to produce consistent results from the nanoscale to the millimetric scale. To the best of our knowledge there are only a few attempts calculating dose profiles in a similar way~\cite{kraemerdose}
 Possible extensions of the model that are in progress will be also mentioned.
\section{The macroscopic scale: Energy deposition of the ion in the medium}
 The key quantity of interest for the ionization process, linking the electron production and the propagation of the ion in the medium, is $d\sigma/dW$, the Singly Differentiated Cross Sections (SDCS). As we showed in our previous papers \cite{ourNIMB,ourEPJD}, it is possible to obtain it by a semiempirical formula by Rudd~\cite{Rudd92},similar to the Bethe-Bloch formula for energy loss, with the inclusion of corrections for imposing a correct asymptotic behaviour.:
 \begin{equation}
\frac{d\sigma(W,T)}{dW}= z^2 \sum_i \frac{N_i}{I_i^2}f\left(\frac{W}{I_i},\sqrt{\frac{m}{M}\frac{T}{I_i}}\right)
\label{sdcs}
\end{equation}
Where $f$ is a generic function, with a number of fitted parameters, of $W$, the kinetic energy of emitted electrons and $T$, the incoming projectile energy per nucleon ($E/u$), scaled by $I_i$  the ionization potential of a given shell $i$, with occupation number $N_i$, and where $m/M$ is the ratio between the electron and ion masses, and $z$ the ion's charge(for the details see\cite{ourEPJD}).

 The correct value of $z$, different from the initial bare ion charge, is one of the most important correction to include. One has indeed , to take into account the physical effect of electron capture, experienced by the ion in the last part of its trajectory, when its velocity is relatively small,  that causes the initially bare ion to collect a number of electrons and to reduce its effective charge.
For the total ionization cross section, obtained by  integration on the secondary electron energies
\begin{equation}
\sigma (T)=\int^{\infty}_0 \frac{d\sigma (W,T)}{dW}dW~,
\label{eq4}
\end{equation}

and plotted in Fig.1, we can see that this effective charge, computed for an ion at velocity $\beta c$  according by a formula reported by Barkas~\cite{Barkas63} (inset of fig.1)
\begin{equation}
z_{eff}=z(1-\exp(-125\beta z^{-2/3}))~,
\label{zeff}
\end{equation}
undergoes a drastic decrease till about one half of its initial value, just in the region where the cross section is maximum (Bragg peak), that is the most important for damaging studies.
This suggests experiments that could mimic the Bragg peak scenario by easiest conditions, that is by simply using a half-stripped carbon ($C^{+3}$)  and a collision energy approaching 100 keV.
Another interesting effect related to this charge variation, consists in a  displacement of the maximum of the curve from  0.1 to 0.3 MeV/u, as can be seen in fig.1.
\begin{figure}
  \includegraphics[height=.3\textheight, angle=0]{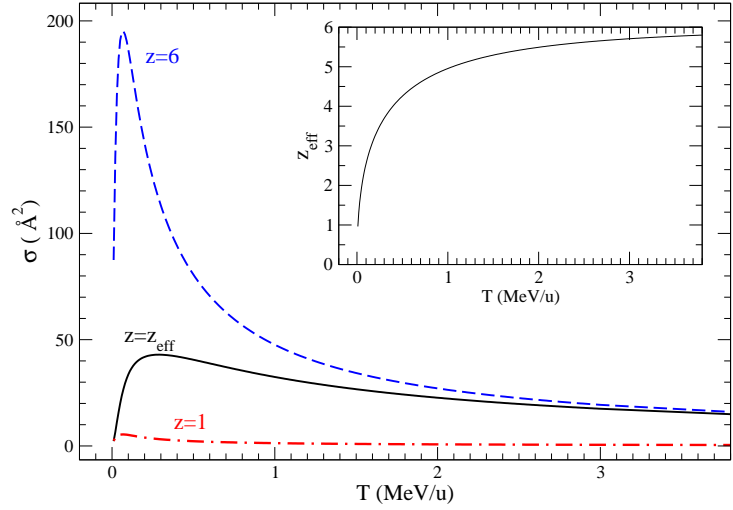}
  \caption{Total Cross section of ionization . In the inset the $z_{eff}$ behaviour as a function of T according to \cite{Barkas63}}
\end{figure}

But the most stringent test of the used formula is, after the integration, to retrieve the Linear Energy Transfer. 
An energy loss by the ion in a single collision, $W+I_i$ ,integrated with the SDCS
and multiplied by the water density $n$ gives the Linear Energy Transfer
(LET, dT/dx) as a a function of the penetration depth  $x$ that shows, on a millimetric scale,the typical Bragg peak profile:
 \begin{eqnarray}
\frac{dT}{dx}=-n\sum\limits_i \int^{\infty}_0
(W+I_i)\frac{d\sigma_i(W,T)}{dW}dW~,
\label{eq6}
\end{eqnarray}
\begin{eqnarray}
x(T)=\int^{T_{0}}_T \frac{dT'}{|dT'/dx|}~,
\label{eq8}
\end{eqnarray}
where $T_0$ is the initial ion energy per nucleon of incidence of the ion.
That allows us to compare (fig.2) with experimental data \cite{Schardt} and simulations with Monte Carlo model for Heavy Ion Therapy (MCHIT)  based on the GEANT4 toolkit \cite{Pshenichnov08}.
 We have already shown in our previous paper\cite{ourEPJD} the importance of the inclusion of a relativistic correction, that shifts the Bragg peak position up to the 40\% of its value. After that the longitudinal deviation in the depth of the Bragg peak is within the 3\% of the whole penetration depth, a value that we can argue to be due to the energy spent for excitation without ionization by the particle. The accounting of this  quantity is presently in progress. The introductionof the effective charge $z_{eff}$ doesn't affect the position of the peak but reduces strongly its height as seen for the $\sigma(T)$. Anyway this reduction is still not sufficient for yelding the correct height and shape of the peak as shown in fig.2 (thin curve), where, for better clearity, the remaining displacement in penetration depth is compensated by properly shifting our curve. 
To this end our calculation does not include the stochastic nature of ion energy loss which appears both in experiments and MCHIT results. Then we introduce such stochasticity by accounting for the energy straggling due to multiple Coulomb scattering of  ion in water \cite{Kundrat07}. A  gaussian spreading of the LET according to a longitudinal-straggling standard deviation parameter $\sigma_{x}=0.8 mm$ computed by Hollmark {\em et al.}~\cite{Hollmark04}, for carbon ions of therapeutic energies, gives an average value
 \begin{eqnarray}
\left\langle \frac{dT}{dx}(x)\right\rangle=\frac{1}{\sigma_{x}\sqrt{2\pi}}\int_{0}^{x_0}\frac{dT}{dx}(x')\exp (-\frac{(x'-x)^{2}}{2\sigma_{x}^{2}})dx'~,  
\label{eqstrg}
\end{eqnarray}
where $x_0$ is a maximum penetration depth of the projectile. With this correction we got perfect agreement with a Monte Carlo simulation where the nuclear fragmentation channels are switched off, that verifies our model in the absence of fragmentation (dashed curves in fig.2).
\begin{figure}
  \includegraphics[height=.35\textheight, angle=0]{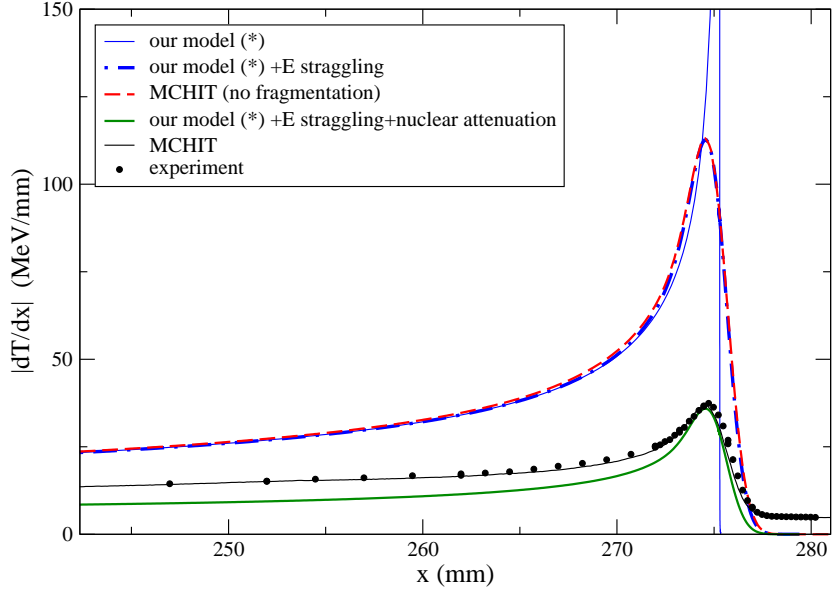}
  \caption{Linear Energy Transfer profiles near the Bragg peak for C$^{6+}$ at 400 MeV/u obtained with different methods. Our data(*) are shifted  by 7.2 mm,in order to compare the shapes}
\end{figure}
We can then simulate the nuclear fragmentation by accounting the fluence attenuation of primary ions due to nuclear reactions along the pathway. The total nuclear charge changing cross section is defined for nuclear fragmentation processes when the initial nucleus of charge Z=6, changes to any other nucleus with Z<6. This value is measurable experimentally and we take the value reported by Chu for incoming energies range consistent with our case) \cite{Chu} : $\sigma_{Z<6}=1248$~mb and  connected to the mean free path
 \begin{equation}
\lambda_{nuc}(T_0)=\frac{1}{n\sigma_{Z<6}}
\end{equation} 
where the mean free path is slowly dependent on the initial energy per nucleon $T_0$, and could be assumed constant in the therapeutical range (from 200 to 450 MeV/u). 
Thus we can consider the  attenuation of the fluence $\Phi_C$  as a factor
\begin{equation}
k(x)=\Phi_C(x)/\Phi_C(0)=exp(-x/\lambda_{nuc}).
\end{equation} 
By including this factor to previous to eq.\ref{eqstrg}, we obtain the effective energy transfer due to primary ions. As evident in figure 2(solid curve), the assumption not to take into account LET components from the nuclear secondary fragments will lead to underestimation in the region before and after the Bragg peak, while the estimation of the Bragg peak height, where the major contribution comes from the specific C ion, is  consistent with experimental data and MCHIT results.

While the analytical method is computationally very fast, our outcome could be used to easily compute also the volumetric energy deposition in the medium, by accounting not only the energy straggling in the propagation direction, but  also in the radial components by an empirical formula provided by Chu depending on the charge and mass of the projectile \cite{Chu} .
\begin{equation}
\sigma_{y}(x;Z,A)=\frac{ax^{b}}{Z^{c}A^{d}}
\end{equation}
where the empirical parameters are respectively  a=0.00373, b=0.896, c=0.207, d=0.396.

It is then possible to map the energy deposition as in figure 3. In that contour plot we can easily see that the scale of this lateral spread due to stochastic effects is of the order of several millimeters. 
  
The experimental procedure usually accounts for a cylindrical spread
of the initial beam with a Full Width at Half Maximum (FWHM) of the order of 3 mm,  that is here also included by an additional lateral spreading term ($\sigma_0=FWHM/(2\sqrt{2ln2})$

\begin{equation}
\left\langle \frac{dT}{dV}(x,y)\right\rangle =\frac{1}{S}\frac{1}{2\pi \sigma_{x}\sqrt{\sigma_{y}^2(x)+\sigma_0^2}}\exp\left(-\frac{y^2}{2(\sigma_{y}^{2}(x)+\sigma_0^2)}\right)\int_{0}^{x_{0}}\frac{dT}{dx}(x')\exp\left(-\frac{(x'-x)^{2}}{2\sigma_{x}^{2}}\right)dx'\label{lateralstr}
\end{equation}
 \begin{figure}
  \includegraphics[height=.3\textheight, angle=0]{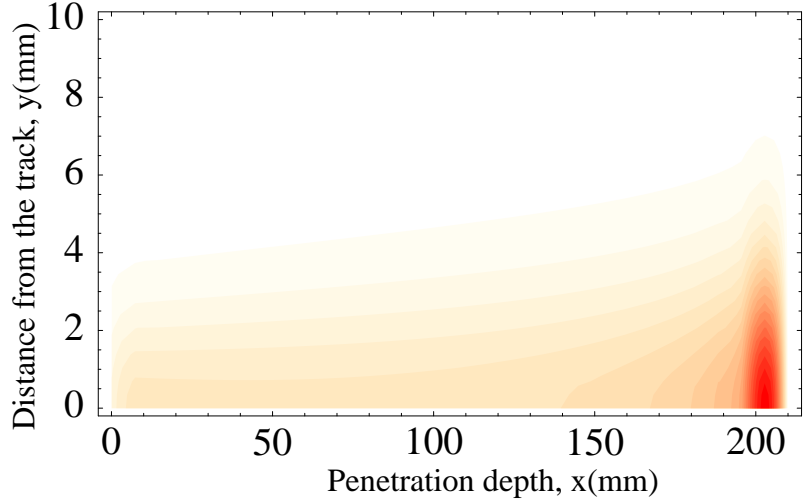}
  \caption{Volumetric energy deposition along and around the track for C$^{6+}$ at 400 MeV/u. Each countour step stays for an increase of 1 MeV/mm$^2$.}
\end{figure}

 where $S$ is the unit of surface for dimensional consistency.
\section{The nanoscopic scale: Secondary Electrons and DNA damage}

Once tested the reliability of the method at large scale, we can extract from the cross sections results information on the damaging electrons that are produced.
 First of all we want to investigate whether the electron showers produced by different single ions have negligible overlap between each other. As it is known from Monte Carlo simulations~\cite{Meesungnoen02} and experimental data~\cite{Watt96}  the penetration lengths of the electrons of that energy range(1--100 eV, as shown below) in water are within 10 nm, we can take as the effective radius of the track, $r_a$, this value. Thus considering a typical ion flux used as $I=10^8$ions cm$^{-2}$s$^{-1}$  and the whole irradiation time, usually 20 min, we get the average number of ions passing an area of the size of the effective track as 
\begin{eqnarray}
 \bar{N}=\pi r_a^2It
\label{poiss}
\end{eqnarray}
and assuming a stochastic process, governed by a Poisson statistics we get the probability of $k$ hits in the same area as 

\begin{eqnarray}
P(k)=e^{\bar{N}}\frac{\bar{N}^{k}}{k!}
\label{eq3}
\end{eqnarray}
that provides already for k=2, P(2)=5$\times$10$^{-2}$, when considering the total irradiation time and P(2)= 1.8$\times$ 10$^{-4}$ when taking t=1 min, time sufficient for repairing mechanisms to start and then levelling out the previous consequences of the damage. We can then consider the damaging effects connected to different tracks to be independent.
 
 One can estimate the damage produced by  ion tracks by considering values plotted in  figure 1. In the Bragg peak region $\sigma_{BP}\approx\sigma(0.3)\approx 40$\AA$^2$, then the number of electrons produced for 1 nm of the ion's trajectory will be (multiplying by the density of water $n=33.4$ nm$^{-3}$) equal to $n\sigma_{BP}\approx13$ nm$^{-1}$.
This quantity divided by the effective area along the track, as mentioned above, gives the number density of the electron plasma along the track:
  \begin{equation}
n_e = \frac{n \sigma_{BP}}{\pi r_a^2} \approx 0.045{\rm nm}^{-3}.
\end{equation}
As it is known the conventional measure of the DNA damage is related to the number of double strand breaks, that is the occurrence of 2 damaging events on opposite strands within a single DNA convolution of length $l_{DNAconv}$=3.4 nm and radius $r_{DNA}$=1.1 nm. Now, considering the volume of that convolution $V_{DNA conv}=\pi r_{DNA}^2l_{DNAconv}=12.9$ nm$^3$ we get the average number of electrons in such a volume, within the effective volume around the track, as $\bar{n}_{e/DNA}=V_{DNA conv}n_e=0.6$. This number reveals that the electron density along the track is on average remarkably high compared to the minimum density requested to have a DSB.
 We can replace this average number by a distribution of the radial density of this electron plasma as a function of the distance from the track, still expressed in units of volume of a DNA convolution, by imposing a gaussian attenuation from the track as
\begin{equation}
n_{e/DNA}(r)=\frac{2N_a}{\sigma_a\sqrt{2\pi}}exp{-(r^2/2\sigma_a^2)}
\end{equation}   
with $\sigma_a=r_a/3$ and $N_a=r_a\bar{n}_{e/DNA}$ and we can see from this profile (figure 4) that at short distances the density is close to 2. This very high density, is a peculiar effect of carbon ions, while e.g. for protons, as mentioned above, the much smaller cross sections will give much lower density. So in this case the simultaneous damage due to different electrons on the same convolution, will be emphasized, while lower densities could produce a double strand break only by the action of the same single electron. That is another reason why we are going to focus our interest on this special kind of damage.  Heavier ions, of course, with their larger charge  would produce even larger densities, but their use shows different contraindications \cite{Kraft05}, in particular related to increased nuclear fragmentation.
\begin{figure}
  \includegraphics[height=.25\textheight, angle=0]{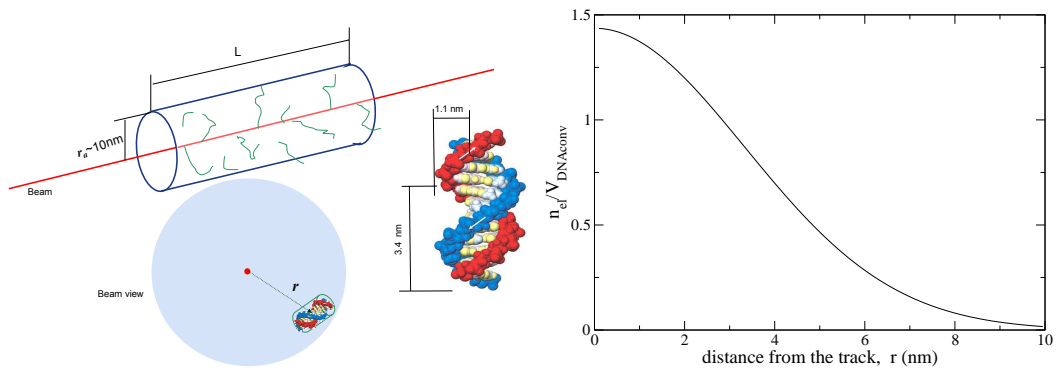}
  \caption{Number density distribution of the electron produced by a carbon ion track for a unit volume equal to a single DNA convolution. On the left, a scheme of the area around the track wandered by the secondary electron, and the DNA convoltion taken as unit.}
\end{figure}
\begin{figure}
  \includegraphics[height=.25\textheight, angle=0]{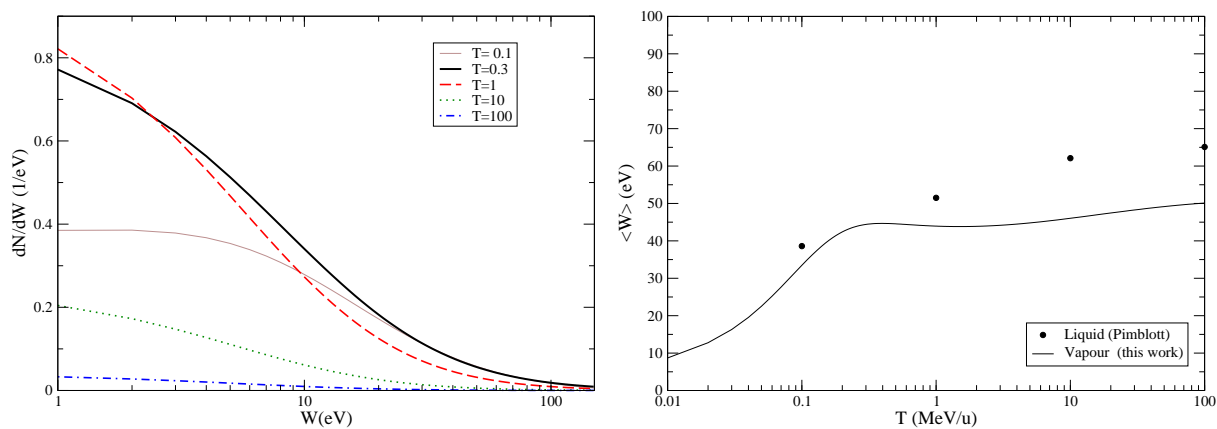}
  \caption{(left panel) Energy distribution of secondary electrons produced by a C$^{6+}$ ion, in a track length of 1 nm, for different values of $T$ (MeV/u). (right panel) Average kinetic energy of produced electrons as a function of the ion energy compared to a similar calculation for liquid water \cite{Pimb07}}.
\end{figure}

 For quantitative estimations, one has to take into account the different trajectories of the produced electrons.  From the SDCS of eq.(\ref{sdcs}) we can extract the spectrum of the energies of the produced secondary electrons in a given $\Delta x$ segment
\begin{eqnarray}
 \frac{dN(W,x)}{dW}= n\Delta x \frac{d\sigma (W,x)}{dW}~.
\label{secel}
\end{eqnarray}
We show in figure 5 (left panel) these data for $\Delta x=$1 nm , and one can clearly see that, at any ion energy considered, the number of produced electrons with W>100 eV is negligible, and a large majority is produced  below 30eV. This result is confirming the importance of the studies of biological effects induced by low energy electrons  that are recently in progress both by experimental \cite{Sanche2000,Sanche2003} and theoretical~\cite{BaccarelliEPJD, Tonzani06} means.

In the right panel of figure 5 it is shown the average energy of produced electrons according to the formula
\begin{eqnarray}
\left\langle W(T)\right\rangle=\frac{1}{\sigma (T)}\int^{\infty}_0 W\frac{d\sigma
(W,T)}{dW}dW~.
\label{eq5}
\end{eqnarray}
The behaviour of this quantity is characterized by a plateau around the Bragg peak, while at high $T$ it reach a value around 60 eV.  The results obtained with different method for the same process but considering liquid water \cite{Pimb07} are also shown for comparison. Even if some deviations are evident, a qualitative agreement in the overall behaviour is achieved.

The obtained electron distribution, dN/dW, can be profitably used for
building  models for quantitative estimation of DNA double strand
breaks induced directly by these electrons or indirectly by radicals creation~\cite{ourRADAM1}.

This approach shows the possibility, even with the use of many simplifications, to face the problem of the damage induced by ion beams in a quantitative way, by including, one by one, all the relevant physical effects appearing at different scales of space and time. We can point out then that there is plenty of room for experiments that may improve the understanding of any single part of these physical processes.




\begin{theacknowledgments}
  The authors are grateful to O. I. Obolensky for fruitful discussions. The financial support of the European Commission within the Network of Excellence project EXCELL is gratefully acknowledged
\end{theacknowledgments}



\bibliographystyle{aipproc}   



\begin{thebibliography}{9}
\bibitem{Kraft05} U. Amaldi, G. Kraft, {\em Rep. Prog. Phys.} {\bf 68}, 1861 (2005).




\bibitem{Goodhead06} D. T. Goodhead, {\em Rad. Prot. Dosim.} {\bf 122}, 13 (2006
).

\bibitem{Nikjoo99} H. Nikjoo, P. O'Neill, M. Terrisol, D. T. Goodhead,
{\em Radiat. Environ. Biophys.} {\bf 38}, 31 (1999)

\bibitem{Nikjoo06} H. Nikjoo, S. Uehara, D. Emfietzoglou, F. A. Cucinotta,
{\em Radiat. Meas.} {\bf 41}, 1052 (2006).

\bibitem{ourRADAM1}A. V. Solov'yov, E. Surdutovich, E. Scifoni, I. Mishustin and W. Greiner, {\em submitted to} RADAM2008 proc.



\bibitem{kraemerdose} M. Kraemer, GSI Report 2000

\bibitem{ourNIMB} O.I. Obolensky, E. Surdutovich, I. Pshenichnov,
I. Mishustin, A. V. Solov'yov, and W. Greiner, Nucl. Inst. Meth. B, {\bf 266}, 1623 (2008)

\bibitem{ourEPJD} E. Surdutovich, O.I. Obolensky, E. Scifoni, I. Pshenichnov, I. Mishustin, A.V. Solov'yov, and W. Greiner, Eur. Phys. J. D, {\em submitted}.


\bibitem{Rudd92} M. E. Rudd, Y.-K. Kim, D. H. Madison, T. Gay,
{\em Rev. Mod. Phys.} {\bf 64}, 441  (1992).

\bibitem{Barkas63} W. H. Barkas, {\em Nuclear Research Emulsions I. Techniques
and Theory}, Academic Press Inc., New York, London, 1963,
Vol. 1, 371.


\bibitem{Schardt} E. Haettner, H. Iwase, D. Schardt, {\em
 Rad. Protec. Dosim.} {\bf 122}, 485 (2006).


\bibitem{Pshenichnov08} I.~Pshenichnov, I.~Mishustin, W.~Greiner,
Nucl. Inst. Meth. B, {\bf 266}, 1094 (2008).


\bibitem{Kundrat07} P. Kundrat, {\em Phys. Med. Biol.} {\bf 52}, 6813 (2007).

\bibitem{Hollmark04} M. Hollmark, J. Uhrdin, D. Belkic, I. Gudowska, A. Brahme, {\em Phys. Med. Biol.} {\bf 49}, 3247 (2004).
\bibitem{Chu}W. T. Chu, B. A. Ludewigt, and T. R. Renner, Rev. Sci. Instrum. 64, 2055 (1993)





\bibitem{Meesungnoen02} J. Meesungnoen, J.-P. Jay-Gerin,
A. Filali-Mouhim, S. Mankhetkorn, {\em Radiat. Res.} {\bf 158}, 657 (2002).
\bibitem{Watt96} D. E. Watt, {\em Quantities for Dosimetry of Ionizing
Radiations in Liquid Water} (Taylor \& Francis, London, 1996).


\bibitem{Sanche2000} B. Boudaiffa, P. Cloutier, D. Hunting,
M. A. Huels, L. Sanche, {\em Science} {\bf 287}, 1658 (2000).

\bibitem{Sanche2003} M. A. Huels, B. Boudaiffa, P. Cloutier,
D. Hunting, L. Sanche, {\em JACS} {\bf 125}, 4467  (2003).

\bibitem{BaccarelliEPJD}I. Baccarelli, F. Sebastianelli, F.A. Gianturco, N. Sanna, {\em Modelling dissociative dynamics of biosystems after metastable electron attachment: the
sugar backbones},
Eur. Phys. J. D     (2008) {\em in press}
\bibitem{Tonzani06} S. Tonzani and C. H. Greene
J. Chem. Phys., {\bf 125}, 094504 (2006) 

\bibitem{Pimb07}S.M. Pimblott, J.A. LaVerne, Rad. Phys. Chem. {\bf 76}, 1244 (2007). 



\end{thebibliography}

\IfFileExists{\jobname.bbl}{}
 {\typeout{}
  \typeout{******************************************}
  \typeout{** Please run "bibtex \jobname" to optain}
  \typeout{** the bibliography and then re-run LaTeX}
  \typeout{** twice to fix the references!}
  \typeout{******************************************}
  \typeout{}
 }


\end{document}